\documentclass[%
 aip,
 amsmath,amssymb,
 reprint,%
]{revtex4-1}

\usepackage{graphicx}
\usepackage{dcolumn}
\usepackage{bm}
\usepackage{color} 

\begin{document}

\preprint{AIP/123-QED}

\title
{{Spin-dependent electron transport in waveguide with continuous shape}}

\author{Yue Ban}

\affiliation{Departamento de Qu\'{i}mica F\'{i}sica, UPV-EHU, Apdo 644, 48080 Bilbao, Spain}

\author{E. Ya. Sherman}
\affiliation{Departamento de Qu\'{i}mica F\'{i}sica, UPV-EHU, Apdo 644, 48080 Bilbao, Spain}
\affiliation{IKERBASQUE, Basque Foundation for Science, 48011, Bilbao, Spain}

\date{\today}

\begin{abstract}

We study effects of the shape of a two-dimensional waveguide on the spin-dependent electron transport in the
presence of spin-orbit coupling. The transition from classical motion to the tunneling regime
can be controlled there by modulating the strength of spin-orbit coupling if the waveguide
has a constriction. The spin precession strongly depends on the shape of the waveguide.

\end{abstract}

\maketitle

The spin-dependent transport of ballistic electrons \cite{Datta-Das}
and the spin manipulation in nanostructures are the core components of spintronics.
Two main mechanisms of spin-orbit coupling (SOC), the structure inversion asymmetry
described by the Rashba term \cite{Rashba} and the bulk
inversion asymmetry described by the Dresselhaus term \cite{Dresselhuas} are
the key elements in these studies. Electron transmission
in a variety of nanostructures including quantum wire with short-ranged irregularities at the boundary \cite{Kunze},
waveguides containing impurities \cite{Lee} or attached to a cavity or a quantum dot \cite{Na_Lee2}
was studied thoroughly
for designing mesoscopic devices \cite{Eugster_Sols}.
It was shown recently that tuning of the gate voltage in a
symmetric quantum point contact can form topological edge states with  spin
polarization robust against the local potential scattering \cite{Chang}.
The fact that the shape of a nanostructure determines the
wavefunctions of the carriers localized there \cite{He}, and as a result,
is responsible for its charge and spin transport properties, motivates the investigation of
transport in  waveguides with a nontrivial shape.

Electron propagation in nonuniform structures in the presence of SOC provides a promising mean to
generate and manipulate the spin-polarized electrons \cite{Zhai,Sanchez-Serra,Serra,Wang,Akgue2,Chang1}.
For example, an electron stub waveguide with SOC was proposed to realize spin filtering and accumulation \cite{Zhai}
and the importance of the lateral geometry and pattern of the static \cite{Bellucci_Xiao}
and time-dependent \cite{Liang} SOC are well understood. The spin scattering with an effective attractive potential
produced by the Rashba interaction \cite{Sanchez-Serra} and with periodically modulated strength of
the spin-orbit interaction \cite{Wang,Krstajic_Malard} were also addressed.
The concept analysis of this type of devices can be found in Ref. \onlinecite{Ting}.

In this Letter, we propose a theory of spin transport in a
quasi-one dimensional semiconductor waveguide with a continuous constriction\cite{Chesi}, and
the Dresselhaus and Rashba SOCs\cite{Chang2}.
By combining these two factors, geometrical shape and the SOC,
we show the effect on the scattering induced by the shape variation and
explore the spin polarization by tuning Rashba coupling strength which can be well controlled
by applying a bias across the structure \cite{Nitta-Koga-Harley}.
It is also shown that the spin states of the transmitted electrons are determined by
the coordinate-dependent SOC. Once designed and produced, this structure has
well-documented measured properties and can be used as a hardware element for a spin transport
device.

\begin{figure}[ht]
\begin{center}
\scalebox{0.55}[0.55]{\includegraphics{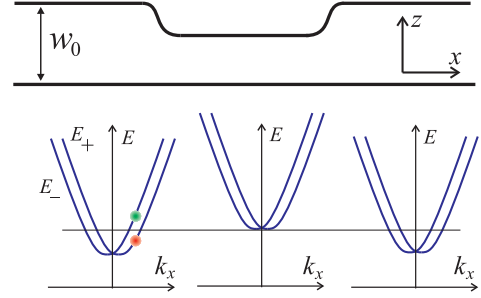}}
\caption{(Color online) Schematic diagram of waveguide with continuous shape, which
experiences the constriction around $x=0$
and tends to
a constant $w_0$ asymptotically.
The corresponding spin-split energy bands are shown in the lower panel,
demonstrating that the effective potential is strongly spin-dependent.
}
\label{model}
\end{center}
\end{figure}

We consider a two-dimensional waveguide [001] grown sketched in Fig. \ref{model}, whose width in the $z$-direction is described by
the function of $w(x)=(\tanh[(x-a)/L]-\tanh[(x+a)/L])w_{1} + w_{0}$, where $a$, $L$, $w_{1}$, and
$w_0$ are the geometry parameters.
The waveguide experiences a continuous change
in the width, with $w(|x|\rightarrow\infty)=w_0$. We assume the boundary conditions, where the wavefunction vanishes at the boundaries. As a result,
the electron propagates along the $x$-axis and is confined in the $z$-direction. In the
$y$-direction, we assume that the wave function is a free plane wave $\exp(i k_{y}y)$ with $k_y=0$.
As shown in Fig. \ref{model}, the constriction-like waveguide can be described as a repulsive potential
since the bottom of the first subband is higher than in the outside regions.
The total Hamiltonian in the presence of Rashba and Dresselhaus SOCs is $\hat{H} = \hat{H}_0 + \hat{H}_{R} + \hat{H}_{D}$.
The kinetic part is $\hat{H}_0=\hat{p}^2_x/2m + \hat{p}^2_z/2m $. Rashba term is $\hat{H}_R = \alpha_R \hat{\sigma}_y \hat{k}_x$,
where $\alpha_R$ is the Rashba coupling constant. The bulk Dresselhaus term is
$\hat{H}_D^{\textrm{bulk}} = \alpha [ \hat{\sigma}_x \hat{k}_x (\hat{k}^2_y - \hat{k}_z^2 )
+ \hat{\sigma}_y \hat{k}_y (\hat{k}^2_z - \hat{k}_x^2 ) + \hat{\sigma}_z \hat{k}_z (\hat{k}^2_x - \hat{k}_y^2 )]$
and $\alpha$ is a parameter for SOC. The effect of the Dresselhaus term $\hat{H}_D$ on the electron in the waveguide is
obtained by orientation-dependent averaging the bulk Hamiltonian on the electron wavefunction \cite{DyakonovRashba},
leading, in general, to the orientation-dependent spin transport \cite{Chang2}.
For convenience, we introduce material and structure related
Dresselhaus coefficient as $\alpha_D=\alpha \pi^2/w^2_0$, while the Rashba coupling
can be modulated by applying the bias in the $z$-direction.
For simplicity, we use dimensionless units, choose $\hbar=m=w_{0}=1$ and introduce
constant $r\equiv\alpha_R/\alpha_D$.
The wave function corresponding to the Hamiltonian $\hat{H}$ can be expressed as the superposition of
eigenstates of all size quantization sub-branches ($n=1,2,3...$) in the ``adiabatic'' basis
\begin{eqnarray}
\label{wavefunction} {\bm{\psi}}(x,z)= \sum_n \sqrt{\frac{2}{w(x)}} {\bm{C}_n}(x)
\sin\left(\frac{n \pi z}{w(x)} \right),
\end{eqnarray}
where ${\bm{C}_n}(x)$ are two-component spinors.
However, if the energy of the particle is lower than the minimum of the second quantization subband,
the contribution of the second state $|{\bm{C}}_{2}(x)|^2$ is less than 0.05
for the structures we consider. Therefore, we come to the single
mode description ($n=1$, ${\bm{C}}(x)\equiv{\bm{C}_1(x)}$) if the incident 
energy satisfies this condition, as it will be assumed below.

In the limit of $|x|\rightarrow\infty$ where the width of the waveguide is a constant $w_0$,
Hamiltonian can be simplified as $\hat{H} = \hat{ H}_0 +\left(\alpha_R \hat{\sigma}_y -\alpha_D \hat{\sigma}_x\right)\hat{k}_x$.
We take the spinor $\bm{C}(x)$ at $x\rightarrow-\infty$  as the superposition of
the incident and reflected waves,
\begin{eqnarray}
\label{C in - infinity}
\bm{C}(-\infty) &=& A_{i}^{[+]}\exp(i k_{+} x) \bm{\gamma_{+}} + A_{r}^{[+]}\exp(-i k_{-} x) \bm{\gamma_{+}}
                         \\ \nonumber &+& A_{i}^{[-]}\exp(i k_{-} x ) \bm{\gamma}_{-} + A_{r}^{[-]}
\exp(-i k_{+}x)\bm{\gamma}_{-},
\end{eqnarray}
while at $x\rightarrow\infty$ it is the transmitted waves,
\begin{eqnarray}
\label{C in + infinity}
\bm{C}(+\infty) = A_{t}^{[+]} \exp(i k_{+} x) \bm{\gamma}_{+} + A_{t}^{[-]}\exp(i k_{-} x) \bm{\gamma}_{-},
\end{eqnarray}
where the spinors $\bm{\gamma}_{+} $ and $\bm{\gamma}_{-}$ are the two eigenstates parallel and antiparallel to the direction of the effective spin-orbit field,
${\bm{\gamma}}_{\pm} = 1/\sqrt{2}\left[\mp\left(1 + i r\right)/\sqrt{1+r^2}, 1\right]^{T},$
and $A_i$, $A_r$ and $A_t$ represent the amplitude of incident, reflected, and transmitted waves,
respectively. Due to the SOC,
the energy band is split into two subbands $"+"$ and $"-"$ one,
corresponding to the spinors $\bm{\gamma}_{+}$ and $\bm{\gamma}_{-}$ respectively.
Corresponding energies $E_{+}$ and $E_{-}$  can be expressed at the incidence momentum $k$ as:
\begin{eqnarray}
\label{E} E_{\pm} = \frac{k^2}{2}+\frac{\pi^2}{2 w_0^2} \pm \alpha_D \sqrt{ 1 + r^2}k.
\end{eqnarray}
The spin splitting $\Delta E = 2 \alpha_D \sqrt{ 1 + r^2}k$ can be enlarged by increasing the tunable parameter $r$.

To obtain the wavefunction, we solve the Schr\"{o}dinger equation by using Eq. (\ref{wavefunction}) and total Hamiltonian. By integrating over $0<z<w(x)$ to eliminate $z-$dependence, one can rewrite the Schr\"{o}dinger equation as: $\bm{C}''+\bm{K}_1\bm{C}'+\bm{K}_2\bm{C}
                +\bm{S}_1\bm{C}'+\bm{S}_2\bm{C}=0 $,
where the matrices are
\begin{eqnarray}
\label{K1 and K2}
\bm{K}_1=\left[\begin{array}{cc} M & 0 \\ 0 & M \end{array}\right],
~~ \bm{K}_2=\left[\begin{array}{cc} G & 0 \\ 0 & G \end{array}\right]\;, \label{B}
\end{eqnarray}
\begin{eqnarray}
\label{S1}  \bm{S}_1 = 2 \alpha_D \left[\begin{array}{cc} -i B w' & - i B+ r \\ -i B - r & i B w'\end{array}\right] ,
\end{eqnarray}
\begin{eqnarray}
\bm{S}_2=\alpha_D \left[\begin{array}{cc} i B (3 M w'- w'') & M (3 i B + r ) \\
\\ M (3 i B - r ) &  -i B (3 M w'- w'')  \end{array}\right]. \label{S_2}
\end{eqnarray}
For brevity, we use notations $w\equiv w(x)$, $w' \equiv w'(x)$, $w'' \equiv w''(x)$, $M \equiv M(x)=w'/w$,
$G \equiv G(x)=2 E -\pi^2/w^2 - M^2(1 + 2 \pi^2 )/4 +w''/2w$, and $B \equiv B(x)=w^2_0/w^2$.
The matrices $\bm{S}_1$ and $\bm{S}_2$ determine the spin evolution due to the coupling between the spin states and the shape of the waveguide.

In the constriction, as shown in Fig. \ref{model}, the ground state energy increases
so that the classical motion or the tunneling, as determined by the
spin-dependent electron energy  can occur.
The minimal energy in the constriction, expressed by following Eq.(\ref{E_b}),
is the threshold energy to open the classically propagating mode in it:
\begin{eqnarray}
\label{E_b} E_b \approx \frac{\pi^2}{2 w^2(0)}- \alpha_D^2 \frac{r^2 + B^2(0)}{2}.
\end{eqnarray}
By tuning the Rashba coupling $r\alpha_{D}$, one can decrease $E_{-}$  from the value
larger than $E_{b}$ to a smaller one, realizing transition from the classical motion to the tunneling.
Figure \ref{E-Transmission}(a) shows the dependence of the split energies $E_{-}$, $E_{+}$, and the minimum of the energy band
in the constriction $E_{b}$ on the controllable ratio $r$, provided by the same incidence momentum $k$. The crossing point of
$E_{-}$  and $E_{b}$ separates the motion through the potential into the classical (left) and the tunneling (right
from the point). This observation opens a possibility to manipulate electron transmission
in the waveguide by a transverse electric
field, solely modulating the SOC. The feasibility of this approach is seen in
Figs. \ref{E-Transmission}(b-d), demonstrating the spin-dependent
transmission $T_{\delta \delta'}$ versus $r$, where $\delta'$ and $\delta$ represent the incidence and
transmission channels, respectively, for different width $L$ of the boundaries of the
constriction. As one can see in the Figure \ref{E-Transmission}, not only the modulation
in the width, but also the details of the shape are important. As a result, by choosing the shape of the waveguide, one
can achieve its required spin transport properties.
\begin{figure}[ht]\label{E-Transmission}
\scalebox{0.6}[0.6]{
\includegraphics*{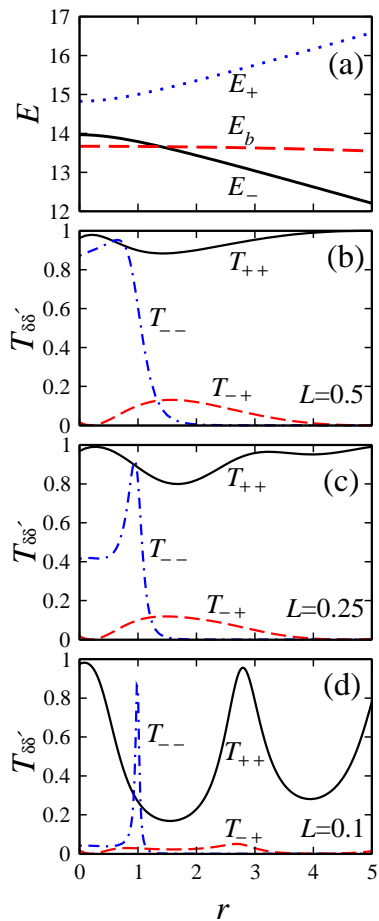}}
\caption{ (Color online) (a) Dependence of $E_{-}$ (solid), $E_{+}$ (dotted) and
$E_b$ (dashed) on $r$.
The waveguide is a repulsive potential, where $a=3$, $L=0.5$, $w_{1}=1/5$ and the strength of SOC is $\alpha=0.01$, as typical for GaAs-based structures.
(b-d) Three main transmissions $T_{\delta \delta'}$ versus $r$ with different boundaries, $L=0.5$ (b), $L=0.25$ (c), $L=0.1$ (d), where $\delta'$ and $\delta$ represent
the incidence and transmission channels. The increase in $T_{--}$ at $r$ close to one is due to the almost resonant tunneling
in this channel at these values of $r$ (cf. Fig. 1).}
\label{E-Transmission}
\end{figure}

\begin{figure}[ht] \scalebox{0.55}[0.55]{
\includegraphics*{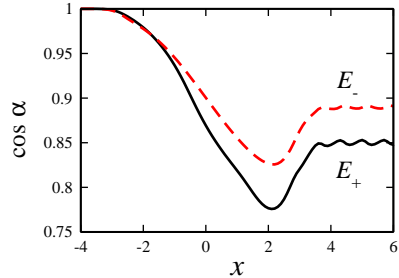}}
\caption{ (Color online) Cosine of the spin angle $\theta$ as a function of $x$ coordinate, with the incidence energies $E_{+}$ (solid line, spin $+$) and $E_{-}$ (dashed line, spin $-$), given by 
$k=4.35$ and 
$r=0.8$. Other parameters are the same as those in Fig. \ref{E-Transmission}(a). }
\label{angle} \end{figure}

To characterize the spin polarization in the transmission, we use the angle $\theta$
between the direction of spin at any $x$, and the one of the initial state at $x\rightarrow-\infty$:
$\cos \theta = {\langle\bm{\sigma}(-\infty)\rangle \cdot \langle\bm{\sigma}(x)\rangle}$,
where $\bm{\sigma}$ consists of the Pauli matrices,
and $\langle\sigma_j(x)\rangle, j=x,y,z$ is the expectation value of spin component at position $x$. Suppose the incident
electrons are in the
given channels, spin ``$+$" (parallel) or ``$-$" (antiparallel) to the spin-orbit field, respectively, with energies in the
form of Eq. (\ref{E}), given by the same incidence momentum $k$. When the electron in any spin channel is propagating in the
constant width region, the angle $\theta$ is zero as the direction of spin keeps unchanged. However, in the region where $w(x)$
starts to decrease, the direction of the spin-orbit field starts to alter due to the change in the Dresselhaus coupling.
As a result, the spin begins to rotate around the axis determined by the local direction of the
spin-orbit field. As the momenta for channel ``$+$" and ``$-$" in the constriction are different, the $\theta$-angles experience
different changes. As shown in Fig. \ref{angle}, the spin angles for the incident electron with spin ``$+$" and ``$-$" begin to separate
at the entrance ($x=-4$) and separate by approximately 0.1 at the exit.

In conclusion, we have investigated the spin transport in a waveguide with a narrowing, whose variation in the width plays
an important role in charge and spin transmission. The transmission can be strongly modified by modulating the magnitude of the
Rashba coupling and can be changed with waveguides oriented along different crystallographic directions. Changeable geometry will provide an alternative way to control the
charge and spin transport. The produced
structures can be applicable to manipulate charge transport by changing electric field across the electron propagation direction
and are promising hardware elements for designing spin transport devices.

This work is supported by the Basque Country Government (IT472-10) and
MICINN (FIS2009-12773-C02-01). Y. B. thanks financial support from the Basque Country Government (BFI-2010-255).

\end{document}